\documentclass{aa}
\usepackage{graphics}
\usepackage{graphicx}
\begin{document}

   \thesaurus{06     
              (03.11.1;  
               16.06.1;  
               19.06.1;  
               19.37.1;  
               19.53.1;  
               19.63.1)} 
   \title{The birth of strange stars and their dynamo-originated
       magnetic fields}


   \author{R.X. Xu
          \inst{1,} \inst{2}
          \and
          F.H. Busse\inst{2}
          }

   \offprints{R.X. Xu}

   \institute{Beijing Astrophysical Center and Astronomy Department,
    Peking University, Beijing 100871, China\\
    email: rxxu@bac.pku.edu.cn
    \and
    Theoretical Physics IV, University of Bayreuth,
    95440 Bayreuth, Germany\\
    email: Friedrich.Busse@uni-bayreuth.de
    }

   \date{Received:~~5 October 2000; ~~~~~Accepted:~~27 March 2001}

   \maketitle

\begin{abstract}

It is shown that protostrange stars (PSSs) can be convective and
that there are two possible scenarios describing their turbulence.
Besides the local turbulence on the scale which is less than the
mean free path of neutrinos, large-scale ($\sim$ 1 km) convection
also may occur with properties that are similar to those of
convection in protoneutron stars (PNSs). We thus suggest that
strange stars can also create dynamo-originated magnetic fields
during the deleptonization episode soon after a supernova
explosion.
Further detailed investigations are needed to see whether or not strange
stars and neutron stars can be distinguished according to the differences
in dynamo actions in strange quark matter and in neutron matter.
The magnetic fields of strange stars and neutron stars may also behave
very differently during the accretion-phase when the fields decay.
%

\keywords{pulsars: general ---
      stars: neutron ---
      elementary particles ---
      supernovae: general ---
      magnetic fields}

\end{abstract}



\section{Introduction}

As ordinary nucleons (neutrons and protons), which are composed of
quarks, are squeezed tightly enough at high temperature and
pressure, they may turn into a soup of deconfined free quarks.
This type of phase transition might be expected in the center of
a neutron star which consists mainly of neutron matter. In the
conventional picture of neutron stars with quark matter, the
neutron (hadron) and quark phases are separated by a sharp
boundary with density discontinuity; such stars are called
``hybrid'' stars. But, as first considered by Glendenning (1992)
for ``complex'' systems, it is possible that bulk quark and
nuclear matter could coexist over macroscopical distances in
neutron stars; such stars are called ``mixed'' stars [see, e.g.,
Heiselberg et al. (2000) for a recent review about hybrid and
mixed stars].
A further radical view is that the {\it whole} neutron star
should be phase-converted to be a {\it strange} star, which
consists of nearly equal numbers of $u$, $d$, and $s$ quarks and
associated electrons for charge neutrality, {\it if} strange
quark matter (SQM) is the true ground state of strong interacting
matter (Bodmer 1971, Witten 1984, see also, e.g., Madsen (1999),
for a recent review of the physics and astrophysics of strange
matter and strange stars).
Strange stars provide {\bf a} sharp contrast to neutron stars
(e.g. the mass-radius relations and the surface conditions),
although both neutron and strange stars may have similar radii
($\sim 10$ km) and masses ($\sim 1.4 M_{\odot}$).

Pulsars can be modeled as neutron stars or strange stars. No {\em
strong} observational evidence known hitherto favors either of
the models. It is thus of great importance and interest in current
astrophysics to distinguish neutron stars and strange stars
observationally.
Recently, at least three hopeful ways have been proposed for
identifying a strange star: 1, hot strange stars (or neutron
stars containing significant quantities of strange matter) may
rotate more rapidly since the higher bulk viscosity (Wang \& Lu
1984, Madsen 1992) of strange matter can effectively damp away the
$r-$mode instability (Madsen 1998); 2, the approximate
mass-radius ($M-R$) relations of strange stars ($M\propto R^3$)
are in striking contrast to those of neutron stars ($M\propto
R^{-3}$). Comparisons of observation-determined relations with
theoretical relations may thus determine whether an object is a
neutron star or a strange star (Li et al. 1999); 3, It is
possible to distinguish ``bare'' polar cap strange stars from
neutron stars via pulsar magnetospheric and polar radiation
because of the striking differences between the polar surface
conditions of the two types of stars (e.g., clear drifting
pattern from ``antipulsars'', Xu, Qiao \& Zhang 1999, Xu, Zhang
\& Qiao 2001).
In this paper, we shall study possible differences in the
processes of magnetic field generation by dynamo action in
strange stars and in neutron stars which could eventually lead to
observational distinctions between the two types of stars.

Magnetic fields play a key role in pulsar life. Unfortunately,
there is still no consensus on the physical origin of strong
fields. Naively, it is supposed that pulsars' fields are the
result of the conservation of magnetic flux during supernova
collapses. This idea faces at least two problems (Thompson \&
Duncan 1993, here after TD93): 1, dynamo action can occur in the
convective episodes of both the iron core and the newborn star; 2,
very strong fields ($>10^{12}$ G) of pulsars are probably not
``fossil fields'' since only a few percent of white dwarfs have
fields in excess of $10^6$ G.
Based on the Newtonian scalar virial theorem, one can estimate the
limiting interior magnetic field
$B_{\rm max}\sim 10^{18} M_1 R_6^{-2}$ G
of a star with mass $M=M_1~M_\odot$ and radius $R=R_6~10^6$ cm
(Lai \& Shapiro 1991).
Both the estimate based on flux conservation and that based on
the virial argument, do not depend on the detailed fluid
properties of the stellar interiors.
Thompson \& Duncan (TD93, Duncan \& Thompson 1992) have
extensively considered turbulent dynamo amplification in
protoneutron stars (PNSs) and in the progenitor stars, and found
that multipolar structure fields as strong as $10^{16}$ G
(``magnetar'') can be generated by PNS convection. This suggestion
seems to have been confirmed by recent observations (e.g.,
Kouveliotou et al. 1998). Furthermore, there arises a possible
way to distinguish strange stars from neutron stars since the
dynamo-originated fields depend on the detailed fluid properties
of the stellar interiors.
One interesting question thus is: Can strange stars create
magnetic fields as strong as $10^{16}$ G? (i.e., can strange
stars act as magnetars?). If strange stars can, is there any
difference between the processes of field generation in PNSs and
that in ``proto-'' strange stars (PSSs)? Theoretically, there is
little work known hitherto on the generation of the magnetic
fields of strange stars. It is supposed that spontaneous
magnetization could result in the generation of a compact quark
star (Tatsumi 2000). Alternatively, we suggest that
dynamo-amplification of the magnetic field could also play an
important role in PSSs with high temperatures.
In this work, we are trying to find whether there are any
differences between the dynamo-originated fields of neutron stars
and that of strange stars, in order to contribute to the debate on
the existence of strange stars.
In our discussion, we presume that the strong magnetic fields of
strange stars originate in the strange cores, rather than in the
crusts with mass $\sim 10^{-5} M_\odot$, since strange stars
produced during supernova explosions cannot have such crusts (Xu
et al. 2001).

One difficult issue in the study of dynamo-originated fields of
strange stars is to determine whether color superconductivity
(CSC) occurs in PSS since the dynamo mechanism may not work
effectively in a superconducting plasma\footnote{
In case of CSC, the field creation and evolution may also be
influenced significantly if there exist {\em fast} dynamos with
magnetic Reynolds number ${\cal R}_{\rm m} \rightarrow \infty$ in
PSSs. }.
Recent calculations, based on a model where quarks interact via a
point-like four-fermion interaction, showed that the energy gap
$\Delta$ of zero-temperature strange matter could be 10-100 MeV
for plausible values of the coupling (Alford et al. 1999); thus
the critical temperature of forming possible quark Cooper pairs
$T_{\rm c}\sim \Delta/2 \sim 5-50$ MeV.
However, $T_{\rm c}$ could be altered if a more realistic
quark-quark interaction is used and/or if the the trapping of
neutrinos in PSSs is included. It is also hard to sufficiently
determine the temperature $T({\rm PSS})$ of PSSs. We therefore
simply assume in this paper that $T({\rm PSS})>T_{\rm c}$ in the
first few seconds (the time-scale for dynamo action of PSS) after
PSS formation.


\section{The birth of strange stars: differential rotation
         \& convection?}

The time-scale of PNSs (or PSSs) is of the order of a few seconds,
which is three orders of magnitude longer than the dynamical one.
It is thus worth studying the dynamical bulk evolution in these
stars, such as differential rotation and convection.
Unfortunately, no numerical model of supernova explosion known
hitherto has included the conversion from PNSs to PSSs, although
such a conversion may help in modelling the burst process [e.g.,
1, helping to solve the present energetic difficulties in getting
type II supernova explosions (Benvenuto \& Horvath 1989); 2,
giving a reasonable explanation for the second peak of neutrino
emission in SN 1987A (Benvenuto \& Lugones 1999)]. The absence of
a numerical model is caused by the lack of (1) a full theory
determining the conditions at which the quark matter phase
transition occurs and of (2) a detailed understanding of the
complex burning process of neutron matter into strange matter.

Nevertheless, some efforts have been made in trying
to understand the transition and combustion processes in detail.
The phase conversion occurs in two steps: first neutron matter
deconfinement occurs on a strong interaction time scale $\sim
10^{-23}$ seconds, then chemical equilibration of the deconfined
quark matter takes place on a weak interaction time scale $\sim
10^{-8}$ seconds. Additional neutrinos and energy are produced in
the second step (Dai, Peng \& Lu 1995). Further calculations of
such transitions (Anand et al. 1997; Lugones \& Benvenuto 1998)
have also considered the effect of strong interactions, the
effect of finite temperature and strange quark mass, and the
effect of trapped neutrinos. Recently, Benvenuto \& Lugones
(1999) explored the occurrence of deconfinement transition in a
PNS modeled by Keil \& Janka (1995) and found that the
deconfinement appears as long as the bag constant ${\check
B}\leq$ 126 MeV fm$^{-3}$. Various estimates of the bag constant
indicate that the preferred value of ${\check B}$ lies in the
range of $60 {\rm MeV ~fm^{-3}}\leq {\check B} \leq 110 {\rm
MeV~fm^{-3}}$ (Drago 1999), which means that deconfinement is
very likely to happen. Such 2-flavor quark matter may be
transformed immediately into a 3-flavor one if SQM is absolutely
stable.
From a kinetic point of view, Olinto (1987) has calculated
for the first time
 the conversion of neutron stars into strange stars, suggesting
a deflagration mode with a burning velocity range from $10^4$ km/s
to a few cm/s. However it is found (Horvath \& Benvenuto 1988;
Benvenuto, Horvath \& Vucetich 1989; Benvenuto \& Horvath 1989)
that such slow modes are unstable. This instability would be
self-accelerated, and the burning should occur finally in
detonation modes, although the transition from deflagration to
detonation has not been well understood (Lugones, Benvenuto \&
Vucetich 1994).
In order for the combustion to be exothermic, there exists a
minimum density $\rho_{\rm c}$ for detonation to be possible,
$\rho_{\rm c} \sim 2$ times of nuclear density for some cases
(Benvenuto \& Horvath 1989). Thus the detonation flame can not
reach the edge of the compact core, and the outer part of the PNS
would be expelled. As a result, a strange star is formed with an
almost ``bare'' quark surface which is essential to solve
completely the ``binding energy'' problem in some current pulsar
emission models (Xu et al. 1999, Xu et al. 2001). Both the extra
neutrino emissivity and the detonation wave can favour a
successful core-collapse supernova explosion.

\subsection{Differential rotation}

It is expected that PNSs (thus PSSs) have a strong differential
rotation (e.g., Janka \& M\"onchmeyer 1989, Goussard et al. 1998).
This issue is uncertain, however, due to the lack of a rotating
core model in the pre-supernova evolution simulations. The iron
core collapses, being triggered by electron capture (for lower
entropy core) and/or photodisintegration (for higher entropy
core), almost in the same way as for free fall. Many initial core
models of highly evolved massive stars without rotation have
appeared in the literature (e.g., Arnett 1977, Bruenn 1985), in
which the density $\rho_0$ - radius $r_0$ relation can be fitted
by
\begin{equation}
\rho_0(r_0) = \sum_{\rm i=1}^{n}\rho_{0i} e^{-\alpha_{\rm i} r_0},
\label{rho}
\end{equation}
where $\rho_{\rm 0i}, \alpha_{\rm i}$ are constants.
For the first order approximation, we let $n=1$ in our following
discussion.
In this case, the Lagrangian mass coordinate ${\cal M}(r_0)$ is
\begin{equation}
{\cal M}(r_0) = {4\pi \rho_{01}\over \alpha_1^3}
       [2 - (2+2\alpha_1r_0+\alpha_1^2r_0^2)
          e^{-\alpha_1r_0}].
\label{cM}
\end{equation}
Setting ${\cal M}(10^8)\sim 1.4M_\odot$ and $\rho_0(10^8)\sim 10^8$
g cm$^{-3}$ for the core with radius $R_0=10^8$ cm, one obtains\footnote{
The density model $\rho_0(r_0) = \rho_{01} e^{-\alpha_1 r_0}$ with
$\rho_{01}=2\times 10^{10}$ g cm$^{-3}$, $\alpha_1=5.4\times 10^{-8}$
cm$^{-1}$, based on which we will calculate the nature of
differential rotation of PSSs (or PNSs), represents well the initial
core model of Bruenn (1985, Fig.1, there $R_0\sim 10^8$ cm,
${\cal M}(10^8)\sim 1.4M_\odot$, $\rho_0(0) = 2\times 10^{10}$
g cm$^{-3}$, and $\rho_0(10^8) = 8\times 10^7$ g cm$^{-3}$).
We thus think that the computation according to this simplified density
model can give the correct order of magnitude result for the actual
collapsed rotating core.
}
$\rho_{01}=2\times 10^{10}$ g cm$^{-3}$, $\alpha_1=5.4\times
10^{-8}$ cm$^{-1}$, by Eq.(\ref{rho}) and Eq.(\ref{cM}).
We {\it assume} that the core before collapse with a {\em uniform}
rotation of period $P_0$ can also be approximated by Eq.(1) for
lack of a rotation core model in simulations, and we investigate
below the collapse of the iron core with the above parameters.
Here we also assume that pulsar rotation is mainly the result of
angular momentum conservation during the core collapse process,
rather than of the kick at birth (Spruit \& Phinney
1998)\footnote{
The recent observation of spin-kick alignment of Crab and Vela
pulsars supports this suggestion.
}.

Based on the approximation, the total rotational energy $E_0$ of
the core is
\begin{equation}
E_0  =  {16\pi^3 \rho_{01} \over 3P_0^2}\cdot
{\Im(\alpha_1, R_0)\over \alpha_1^5},
\end{equation}
and the total angular momentum $M_0$ is
\begin{equation}
M_0 = {16\pi^2 \rho_{01}\over 3P_0}\cdot
      {\Im(\alpha_1, R_0)\over \alpha_1^5},
\end{equation}
where the function $\Im (\alpha, R)$ is defined by
\begin{equation}
\Im (\alpha, R) = 24 -
         (24 +24\alpha R+
         12\alpha^2 R^2+
         4\alpha^3 R^3+
         \alpha^4 R^4)e^{-\alpha R}.
\end{equation}
It is found that $\lim_{\alpha \to 0}\Im(\alpha, R)/\alpha^5 =
R^5/5$. Contrary to the case of the core before collapse, newborn
strange stars may be well approximated as objects with
homogeneous density (Alcock et al. 1986). For a strange star with
mass $M=1.4M_\odot$ and radius $R=10^6$ cm, we find the density
$\rho\sim 7\times 10^{14}$ g cm$^{-3}$. During the adiabatic
collapse process, in which the entropy of each mass element does
not change significantly, {\em toroidal} forces are negligible
(although a poloidal force can cause a shock wave). The angular
momentum of each mass element is thus conserved, and PSSs should
be in differential rotation. We assume that the material at a
shell with radius $r_0$ of the iron core before collapse
contracts to another shell with radius $r$ in a PSS,
\begin{equation}
r^3 = {3\over 4\pi \rho}{\cal M}(r_0).
\label{r-r_0}
\end{equation}
According to angular momentum conservation of each mass element,
one arrives at the velocity field $V$ dependent on $r$ and $\theta$ (polar
angle),
\begin{equation}
V = {2\pi\sin\theta \over P_0} \cdot {r_0^2 \over r},
\label{v}
\end{equation}
where $r_0$ is a function of $r$ through Eq.(\ref{r-r_0}), and the
mass conservation law $4\pi r_0^2 \rho_0 {\rm d}r_0=4\pi r^2 \rho
{\rm d}r$ has been used. Owing to the momentum transport by
neutrinos, magnetic fields, and turbulence, a uniform rotation is
approached after a certain time and the final rotational energy
$E_{\rm f}$ can be estimated as
\begin{equation}
E_{\rm f} = {16\pi^3 \over 15P^2}\rho R^5\sim 2.3 \times 10^{46}P^{-2},
\end{equation}
corresponding to the angular momentum $M_{\rm f}$,
\begin{equation}
M_{\rm f} = {16\pi^2 \over 15P}\rho R^5,
\end{equation}
where $P$ is the period of newborn uniformly rotating strange
star. According to angular momentum conservation in the collapse
process, $M_0=M_{\rm f}$, the rotation period $P_0$ of the initial
iron core before collapse is
\begin{equation}
P_0 = {5 \rho_{01} \Im(\alpha_1, R_0)\over \rho R^5 \alpha_1^5} P
\sim 4.68\times 10^3 P,
\label{P0}
\end{equation}
for typical parameters, i.e., the core rotates with a period of
about 50 seconds for an initial pulsar period $P\sim 10$ ms.
From Eq.(\ref{v}), one obtains the velocity derivative $|\nabla
V|$,
\begin{equation}
|\nabla V| = \sqrt{({\partial V\over \partial r})^2
+({1\over r} {\partial V\over \partial \theta})^2},
\label{dv}
\end{equation}
as a function of $\theta$ and $r$.

\begin{figure*}
  \includegraphics[width=7cm,height=7cm]{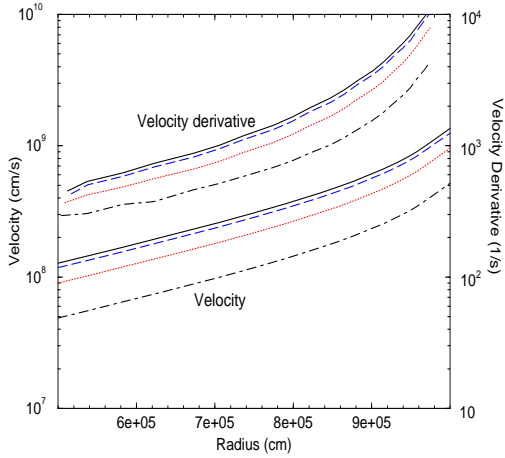}
  \hfill
  \parbox[b]{8cm}{
    \caption{The calculated velocity
and velocity derivative profiles for different polar angles
$\theta$. Dot-dashed, dotted, dashed, and solid lines are for
$\theta=\pi/8, \pi/4, 3\pi/8, \pi/2$, respectively. The lower
(upper) 4 lines, scaled by the left (right) ordinate, are for
velocity (velocity derivative). $P=10$ ms in the calculations.}
    \label{Fig.v_dv}}
  \end{figure*}

Fig.\ref{Fig.v_dv} shows the velocity and the velocity derivative
of this differential rotation scenario for $P=10$ ms,
respectively, based on Eq.(\ref{v}) and Eq.(\ref{dv}). We find
$|\nabla V|\sim 10^4$ s$^{-1}$ in the outer part of PNSs or PSSs.
Such differential rotation may play an important role in the
creation of magnetic fields of PSS or PNS, as will be discussed in
section 3.
%

\begin{figure*}
  \includegraphics[width=7cm,height=6cm]{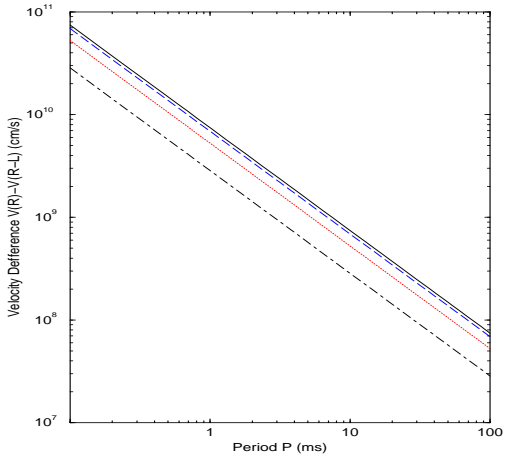}
  \hfill
  \parbox[b]{8cm}{
    \caption{The
velocity difference in the convective outer layer with thickness
of $L\sim 1$ km as a function of pulsar initial period $P$ for
different polar angles $\theta$. See Fig.\ref{Fig.v_dv} for the
definition of lines.}
    \label{Fig.Dv}}
  \end{figure*}

Since the dynamo actions are in the outer convective layer with
thickness $L\sim l_{\rm p}\sim 10^5$ cm (see Eq.(\ref{l_p}) and
section 3), we give the velocity difference of differential
rotation, $\Delta V=V(r=10^6~{\rm cm})-V(r=9\times 10^5~{\rm
cm})$, as a function of $P$ in Fig.{\ref{Fig.Dv}}.
%

\subsection{Convection}

Turbulent convection in PNSs has been extensively investigated
before (e.g. Burrows \& Lattimer 1988, Wilson \& Mayle 1988,
Miralles, Pons \& Urpin 2000). In the Kelvin-Helmholtz cooling
phase, both the negative gradient of the entropy and of the
lepton fraction can drive convection in newborn neutron stars,
given that the mean free paths of leptons are much smaller than
the convection length scale.
Protostrange stars form after the SQM phase-transition with a
time scale of $\sim R/c \sim 10^{-4}$ s, which is much shorter
than that of neutrino diffusion and thermal evolution, assuming
that the actual combustion mode is detonation (Benvenuto, Horvath
\& Vucetich 1989).

We expect that PSSs are convective, since negative gradients of
entropy and neutrino fractions would also appear in the outermost
layers which can lose entropy and neutrinos faster than the inner
part. Following arguments used by Thompson \& Duncan (TD93) in
the case of PNSs, we compare the radiative and adiabatic
temperature profiles to see whether convection occurs in PSSs. We
assume neutrinos in the outer part of PSS are nondegenerate since
there the neutrinos are lost rapidly. Iwamoto (1982) has shown
that for the mean free path $l$ of the scattering of nondegenerate
neutrinos by relativistic degenerate quark matter with
temperature $T$, $1/l\propto T^3$ holds. Thus, the neutrino
opacity in SQM scales as $T^3$ (rather than $T^2$ as for the case
of PNSs), and the radiative temperature profile in PSSs is much
steeper, $T(r) \propto p(r)$. We can employ the analytical
equation of state for SQM in the case of zero strange quark mass
($m_{\rm s}=0$) and zero coupling constant ($\alpha_{\rm s}=0$)
of strong interaction (Cleymans et al. 1986; Benvenuto et al.
1989) to estimate the adiabatic temperature-pressure relation.
The entropy per baryon is $S \sim 3\pi^2T/\mu$ for the case of
chemical potential $\mu\sim 300$ MeV and temperature $T\sim 30$
MeV. Therefore $T(r)\sim (p(r)+{\check B})^{1/4}$ for SQM moving
with fixed $S$ is much less steep, and we can thus expect that a
negative entropy gradient appears in the outer layer which is
unstable to convection, similar to the case of PNSs.
Accordingly we suggest that Schwarzchild convection exists in
PSSs, whereas it will depend on detailed simulations whether
Ledoux convection in PSSs can
be established.
If the timescale $\tau_{\rm NM}$ for neutron matter convection is
much smaller than the neutrino diffusion time (a few seconds), we
may expect that Ledoux convection takes place in a PSS since
the negative gradient of the lepton fraction is nearly the same
for PSSs as that for PNSs.

In the following, we try to estimate the properties of convection
in PSSs. The local pressure scale height $l_{\rm p}$ in PSS is
\begin{equation}
l_{\rm p} = {p\over \rho g} \sim 2 \times 10^5~~{\rm cm},
\label{l_p}
\end{equation}
where $g \sim {GM\over R^2}=1.33\times 10^{14} M_1~R_6^{-2}$ cm
s$^{-2}$ ($M_1=M/M_\odot$, $R_6$ is the radius $R$ in 10$^6$ cm)
is the gravitational acceleration, $p = (\rho c^2 - 4{\check
B})/3\sim 10^{34}$ dyne cm$^{-2}$ is the typical pressure in the
outer convective layer, and the typical density there being
chosen as $\rho = 5 \times 10^{14}$ g cm$^{-3}$ (Alcock et al.
1986). We can assume the thickness $L$ of the outer layer where
dynamo action exists to be equal to this scale height, $L\sim
l_{\rm p}$, and the temperature gradient $\nabla T$ thus is
\begin{equation}
\nabla T \simeq {T\over L} \sim 5 \times 10^5T_{11}~~{\rm K/cm},
\end{equation}
where $T_{11}$ is the SQM temperature in $10^{11}$ K.
The numerical equation of state of SQM including the effect of
non-zero strange quark mass [Eq.(29) of Chamaj \& Slominski 1989]
is
\begin{equation}
\begin{array}{lll}
\rho & = & 1.07 \times 10^{14} {\check B}_{60} + 3{\check P}/c^2 +
       1.02 \times 10^{-9} {\check P}^{0.6455}\\
 & & + 1.34 \times 10^{-14} T^2 {\check P}^{0.09619},
\end{array}
\label{EoS}
\end{equation}
where ${\check P}/c^2\equiv p/c^2 + 1.07 \times 10^{14} {\check B}_{60}$
(${\check B}_{60}$ is the MIT bag constant in 60 MeV fm$^{-3}$, $p$ is
the external pressure).
According to Eq.(\ref{EoS}) the coefficient of thermal expansion
$\alpha$ is
\begin{equation}
\alpha = - {1\over \rho} ({\partial \rho \over \partial T})
       = 2.68 \times 10^{-18} T_{11} {\check P}^{0.09619}\rho_{15}^{-1}
         ~{\rm K}^{-1}.
\label{alpha}
\end{equation}
$\alpha = 1.33\times 10^{-14}~$K$^{-1}$ for $T=10^{11}$K and $p=0$
(Usov 1998).

There are two factors which can cause viscous stresses in a PSS:
neutrino transport and quark scattering.
Neutrino-induced viscosity dominates in a PSS on scales that are
large compared to the neutrino mean-free path $l$. We use the
neutrino mean-free path of nondegenerate neutrino scattering
in SQM for a dimensional estimate (Iwamoto 1982),
\begin{equation}
l = 1.7\times 10^2~\rho_{15}^{-2/3} E_{100}^{-3} ~~{\rm cm},
\label{l}
\end{equation}
where $\rho_{15}$ is SQM density in $10^{15}$ g cm$^{-3}$ and
$E_{100}$ is neutrino energy in 100 MeV. According to Eq.(11) of
TD93, the neutrino mean free path in nuclear matter, $l^{\rm
N}\sim 10^2~\rho_{15}^{-1/3}$ cm for $T=30$ MeV, which is of the
same order as $l$. Unfortunately, no well-determined neutrino
viscosity in PSS has appeared in the literature. We thus just
estimate the neutrino-induced viscosity by a simple kinetic
argument (Wilson \& Mayle 1988)
\begin{equation}
\nu = {1\over 3} lc\xi = 1.7 \times 10^{10}\rho_{15}^{-2/3} E_{100}^{-3}
~~{\rm cm^2~s^{-1}},
\label{nu}
\end{equation}
where we have assumed the ratio $\xi$ of the neutrino energy density
to the quark one to be $\sim 10^{-2}$.
However, the kinematic viscosity due to quark scattering
in SQM (Heiselberg \& Pethick 1993) is much smaller
\begin{equation}
\nu_{\rm q}
\sim 0.1~\rho_{15}^{14/9}~(\alpha_{\rm s}/0.1)^{-5/3}~T_{11}^{-5/3}~~
{\rm cm^2~s^{-1}}.
\end{equation}
Therefore turbulent convection may have a scale $\la l$, which finally
should be damped on a small scale by $\nu_{\rm q}$.

Two possibilities arise for the scenario of turbulence in PSSs.
The first one is that the neutrino viscosity can effectively inhibit
a large-scale convection with length scale $L$, and local convection with
scale $\la l$ will exist.
In this case the local thermal diffusivity due to quark scattering
in SQM (Heiselberg \& Pethick 1993) is
\begin{equation}
\kappa_{\rm q}\sim
0.39~\rho_{15}^{2/3}~(\alpha_{\rm s}/0.1)^{-1}~T_{11}^{-1}~~{\rm cm^2~s^{-1}}.
\end{equation}
The Prandtl number is
\begin{equation}
P_{\rm rq}={\nu_{\rm q}\over \kappa_{\rm q}}\sim 0.25,
\end{equation}
the Rayleigh number is
\begin{equation}
R_{\rm aq}={\alpha g \nabla T l^4\over \kappa_{\rm q}\nu_{\rm q}}
\sim 7.2\times 10^{16},
\end{equation}
and the Coriolis number is
\begin{equation}
\tau_{\rm q}={2\Omega l^2\over \nu_{\rm q}}\sim 2\times 10^8.
\end{equation}
Malkus (1954) had estimated the mean-square values ($v^2$) of the
fluctuating velocity from the Boussinesq form of the hydrodynamic
equations, and found  [Eq.(64) in Malkus (1954)]
\begin{equation}
v \sim {\kappa\over 3d}(R_{\rm a}-R_{\rm ac})^{1/2},
\label{vv}
\end{equation}
where $\kappa$ is the thermal diffusivity, $d$ is the length scale
of convection, $R_{\rm a}$ is the Rayleigh number, and $R_{\rm ac}$
is the critical Rayleigh number.
This relation was supported by experiments [Fig.2 in Malkus (1954)]
for $R_{\rm a}=10^5-10^9$.
Also, Clever \& Busse (1981) found the perturbation energy ($v^2$)
increases nearly proportional to $R_{\rm a}-R_{\rm ac}$ at values
of $R_{\rm a}>10^3$ for low-Prandtl-number convection
[$P_{\rm r}\ga 0.001$, see Fig.11 in Clever \& Busse (1981)].
According to Eq.(\ref{vv}), the turbulent convective velocity on
this scale is
\begin{equation}
v_{\rm q}\sim 3.5\times 10^5~~{\rm cm/s},
\label{vq}
\end{equation}
if we choose $\kappa=\kappa_{\rm q}$, $d=l$, $R_{\rm a}=R_{\rm aq}$.

The second scenario is that the neutrino viscosity is not high enough
to inhibit the large-scale convection, and convection with scale $L$ is
possible.
We can also obtain some dimensionless numbers for this case.
Since both the thermal energy and momentum in PSSs are transported by
neutrinos now, the thermal diffusivity $\kappa$ can be estimated to be
(Wilson \& Mayle 1988)
\begin{equation}
\kappa = {1\over 3} lc = 1.7 \times 10^{12}\rho_{15}^{-2/3} E_{100}^{-3}
~~{\rm cm^2~s^{-1}},
\label{kappa}
\end{equation}
by a simple kinetic argument.
Thus the Prandtl number  is
\begin{equation}
P_{\rm r} = {\nu \over \kappa} = \xi
   \sim 0.1 -0.01.
\label{Pr}
\end{equation}
For a rotating spherical fluid shell with thickness $L$ and
angular velocity $\Omega$, the Rayleigh number is
\begin{equation}
R_{\rm a} = {\alpha g \nabla T L^4\over \kappa\nu}
      \sim  6.0 \times 10^{-23}~~T~L^3.
\label{R}
\end{equation}
i.e. $R_{\rm a}\sim 6.0\times 10^3$ for $T \sim 10^{11}$ K, $p=0$
and $L\sim 10^5$ cm.
The Taylor number $\tau^2$ can be calculated by
\begin{equation}
\tau = {2\Omega~L^2\over \nu}
     = 1.2 \times 10^{-10}~\rho_{15}^{2/3} E_{100}^3~\Omega~L^2.
\label{tau}
\end{equation}
$\tau\sim 1.2\times 10^3$ for $\Omega\sim 10^3$ s$^{-1}$ and
$L\sim 10^5$ cm.
The critical Rayleigh number $R_{\rm ac}$ can be estimated
(Zhang 1995) by
\begin{equation}
R_{\rm ac} \simeq \sqrt{5 \tau}~10^2\sim 10^4.
\end{equation}
Turbulent convection is thus possible for $R_{\rm a}>R_{\rm ac}$
with a velocity which  can be estimated to be ($d=L$)
\begin{equation}
v\la 4\times 10^8~~{\rm cm/s},
\label{vnu}
\end{equation}
according to Eq.(\ref{vv}). Alternatively, if  the mixing-length
prescription (B\"ohm-Vitense 1958, TD92) for a fluid of
semidegenerate fermions is adapted, the convective velocity in
PSS becomes
\begin{equation}
v_{\rm ml} = ({\Gamma-1\over 2\Gamma} {L_{52}\over 4\pi R^2\rho})^{1/3}
\sim 6.8 \times 10^7 L_{52}^{1/3} \rho_{15}^{-1/3}~~{\rm cm~s^{-1}},
\label{vv0}
\end{equation}
where $\Gamma \equiv {\partial {\rm ln}p\over \partial {\rm
ln}\rho} ={\rho c^2 \over \rho c^2 - 4{\check B}}\sim 5$,
$L_{52}$ is the convective luminosity in $10^{52}$ erg s$^{-1}$.
We note that $v_{\rm ml}$ and $v$ are nearly of the same order,
while for PNSs with $\Gamma=5/3$, the convective velocity is $5.4
\times 10^7 L_{52}^{1/3} \rho_{15}^{-1/3}$ cm s$^{-1}$ $\sim
10^8$ cm s$^{-1}$ for $\rho\sim 5\times 10^{14}$ g cm$^{-3}$.
Thus both, the convection in PSSs and in PNSs have convective
velocities of about $10^8$ cm s$^{-1}$ if large-scale convection
is possible. Also the overturn timescales of PSSs and of PNSs are
of the same order, $\tau_{\rm con}=l_{\rm p}/v\sim$ a few ms, and
the Rossby number is $R_{\rm o}=P/\tau_{\rm con}\sim 10^3P_1$ in
this possibility ($P_1$ is the value of the initial period in
seconds).

Another cause of turbulent motion in PSS (and PNS) is differential
rotation.
The Reynolds number ${\cal R}$ of this shear flow is
\begin{equation}
{\cal R} = {\nabla{\check U}{\check L}^2\over \nu},
\end{equation}
where $\nabla{\check U}$ and ${\check L}$ are typical velocity
derivative and length scales, respectively. For the local
convection with scale $\la l$, ${\cal R}_{\rm q}\sim 10^8$. For
the  convection with scale $L$, ${\cal R}\sim 5.8\times 10^3$.
According to the experiment by van Atta (1966, Fig.3 there), turbulence
requires Reynolds number ${\cal R}>10^4$, thus in our case small scale
($\la l$) turbulence may appear owing to the differential rotation.

\section{Dynamo actions in proto-strange stars}

Magnetic fields in a static plasma of finite electric conductivity
$\sigma$ are subject to diffusion and dissipation, the timescale of
which is $\tau_{\rm ohm}={{\check L}^2/\kappa_{\rm m}}
={4\pi\sigma {\check L}^2/c^2}$ ($\kappa_{\rm m}=c^2/(4\pi \sigma)$ is
the magnetic diffusivity).
However, in a flowing plasma, magnetic fields may be created
 with certain velocity fields. This is the so-called
``dynamo'' process, which converts kinetic energy into magnetic
energy (e.g., Moffatt 1978). The equation describing the
time-variation of the magnetic field ${\bf B}$ governs the dynamo
action
\begin{equation}
{\partial {\bf B} \over \partial t} =
\nabla \times ({\bf v}\times {\bf B})+{\cal R}_{\rm m}^{-1}\nabla^2{\bf B},
\label{B}
\end{equation}
where ${\cal R}_{\rm m}={\tau_{\rm ohm}/\tau_{\rm v}}
={\check U} {\check L}/\kappa_{\rm m}$ is the
magnetic Reynolds number ($\tau_{\rm v}={\check L}/{\check U}$ is the
advection timescale). The magnetic diffusivity can be estimated
\begin{equation}
\kappa_{\rm m} = {c^2\over 4\pi \sigma}
    = 5.8 \times 10^{-3} \rho_{15}^{-8/9}~
    (\alpha_{\rm s}/0.1)^{5/3} T_{11}^{5/3}~~
    {\rm cm^2~s^{-1}}
\label{km}
\end{equation}
for SQM [here $\sigma$ is calculated from Eq.(39) and Eq.(28) of
Heiselberg \& Pethick (1993)]. Thus ${\cal R}_{\rm m}\sim
10^{16}$ for PSSs. But for PNSs, the magnetic diffusivity is
(TD93)
\begin{equation}
\kappa_{\rm m}^{\rm N} = 1.5 \times 10^{-5} \rho_{15}^{-1/3}~~
    {\rm cm^2~s^{-1}},
\label{kmN}
\end{equation}
for neutron matter with an electron fraction of 0.2, which is about
two orders smaller than $\kappa_{\rm m}$.
From Eq.(\ref{nu}) and Eq.(\ref{km}), one obtains the magnetic
Prandtl number $P_{\rm m}$
\begin{equation}
P_{\rm m} = {\nu \over \kappa_{\rm m}}
   = 3.0 \times 10^{12}~\rho_{15}^{2/9}E_{100}^{-3}~
     (\alpha_{\rm s}/0.1)^{-5/3}~T_{11}^{-5/3}.
\label{Pm}
\end{equation}

\subsection{An estimate of field strength}

Thompson \& Duncan (TD93) concluded that the dominant kinetic energy to
be converted into magnetic energy in the dynamo action of PNSs is the
convective energy since the initial pulsar rotation period is probably
much larger than the overturn time ($\sim 1$ ms) of a convective cell,
although large-scale $\alpha-\Omega$ dynamo action is essential for
neutron stars with $R_{\rm o}\la 1$ (i.e., initial period $\la 1$ ms)
to produce very high fields (``magnetars'', Duncan \& Thompson 1992).
However, here we suggest that rotation can not be neglected even for
pulsars with typical initial periods, because the differential rotation
energy density
${E_{\rm d}-E_{\rm f}\over R^3} \sim 8\times 10^{31}$ erg cm$^{-3}$
is even {\em larger} than the turbulent energy density
$\rho v^2/2 \sim 5\times 10^{30}$ erg cm$^{-3}$ for $P=10$ ms.
If the large-scale convection scenario discussed in section 2.2
is possible, most of the differential rotation energy may be
converted to magnetic energy by dynamo action.
Let's first estimate the total
differential rotation energy $E_{\rm d}$ in the case when the angular
momentum of each mass element is conserved in the collapse,
\begin{equation}
E_{\rm d} = {16\pi^3\rho\over 3P_0^2}\int_0^R r_0^4 {\rm d}r
\sim 3.1 \times 10^{46}P_1^{-2}~~{\rm erg}.
\label{Ed}
\end{equation}
For a typical initial period $P=10$ ms, one gets $E_{\rm d}\sim
10^{50}$ erg. The dominant energy in the core-collapse supernovae
is the gravitational energy $E_{\rm g}\sim 0.6~GM^2/R \sim 3
\times 10^{53}$ erg for $M=1.4~M_\odot$ and $R=10^6$ cm, if a
strange star is residual. We thus find that a significant part
($\sim$ 0.03\% for $P=10$ ms) of the gravitational energy {\em
has to} be converted to the rotation energy if angular momentum
is conserved.
If most of this differential rotation energy would be converted to
magnetic energy by $\alpha-\Omega$ dynamo action,
${B^2\over 8\pi}\sim {E_{\rm d}-E_{\rm f}\over R^3}$,
one obtains the saturation magnetic field in the interior of the stars,
\begin{equation}
B_{\rm sat}\sim 4.3\times 10^{14} P_1^{-1}~~{\rm G},
\end{equation}
which is near the value estimated by Thompson \& Duncan (TD93) who assumed
that most of the convective energy is converted to the magnetic one.
$B_{\rm sat}\sim 10^{16}$ G for a typical period $P=10$ ms.
Certainly, the actual ``dipole'' magnetic field $B_{\rm p}$ of pulsars
should be only a fraction of $B_{\rm sat}$. For example, in the upper
convection zone of the sun, the rms value the magnetic field is only
$10^{-2} B_{\rm sat}$. If the same is assumed for the pulsars, the rms
value of fields in
PSSs or PNSs should be $\sim 10^{12} P_1^{-1}$ G. However, the poloidal
magnetic field should be another fraction of this rms value, thus
\begin{equation}
B_{\rm p} \la 10^{12} P_1^{-1}~~{\rm G}.
\label{Bp}
\end{equation}
Observations for ordinary pulsars show that the fields are distributed
from $\sim 10^{11}$ G to $5.5 \times 10^{13}$ G (PSR J1814-1744).
Therefore, it is likely the initial period $P$ of ordinary pulsars
could be in the range of a few hundreds to decades of milliseconds.
If $P<10$ ms, much stronger dipole fields (e.g., $\sim 10^{14}-10^{15}$
G for magnetars) can be generated.

In the other scenario, where diffusivities based on neutrino
scattering inhibit large-scale convection, the differential rotation
is damped due to the momentum transport by neutrinos.
In this case the dynamo may be
driven dominantly by turbulent convection, and the field strength
can be estimated as
$B=\sqrt{4\pi\rho v_{\rm q}^2}\sim 10^{13}$ G,
assuming an equilibrium state between kinetic and magnetic energy.
The efficiency of converting differential rotation energy into
magnetic energy by dynamo action in this scenario may be much smaller
than in the case where  convection with the large scale $L$
exists since a significant fraction of differential rotation energy is
likely to be converted into thermal energy due to the high neutrino
viscosity.

Thus, dynamo action may amplify significantly the field before PSSs
cools down to temperatures $T$ smaller than the critical one,
$T_{\rm c}$. When $T<T_{\rm c}$, CSC appears, and the field would exist
as a fossil one for a very long time since $\tau_{\rm ohm}\rightarrow
\infty$ if $\sigma\rightarrow \infty$.
In fact, Alford et al. (2000) investigated recently the effect of CSC
on the magnetic fields, and found that, unlike the conventional
superconductors where weak magnetic fields are expelled by the Meissner
effect, color superconductors can be penetrated by external magnetic
fields and such fields can exist stably on a timescale longer than
the cosmic age.

Eq.(\ref{Bp}) may have an observational consequence for pulsar
statistics, which in turn could test Eq.(\ref{Bp}). In the
magnetic dipole model of pulsars, $(3.2 \times 10^{19})^2 P{{\rm
d}P\over {\rm d}t}=B_{\rm p}^2$ (note: fields are assumed not to
decay in SQM), and the rotation period $P$ is a function of time
$t$ due to energy loss. $P(0)$ denotes the initial period of
pulsars. Considering Eq.(\ref{Bp}), one has $P(t){{\rm
d}P(t)\over {\rm d}t} = \zeta P(0)^{-2}$, the solution of which is
\begin{equation}
P(t)^2=P(0)^2+2\zeta~P(0)^{-2}t,
\end{equation}
where $\zeta \sim 4.3\times 10^{-18}~{\rm s^3}$. Observational
tests of  this equation are highly desirable.

The condition $E_{\rm d}<E_{\rm g}$ gives a limit for the initial
pulsar period $P$. Based on Eq.(\ref{Ed}), one gets $P>0.2$ ms.
Actually, $P\gg 0.2$ ms since the efficiency of converting
gravitation energy into differential rotation energy during
collapse may be very small. It is thus doubtful that supernovae
can produce pulsars with submilliseconds periods.
For recycled millisecond pulsars the above estimate is not relevant.

\subsection{Fast dynamos}

The magnetic field amplification processes in newborn pulsars are
essentially {\em fast} dynamos because of the high magnetic Reynolds
numbers (${\cal R}_{\rm m}\sim 10^{16}-10^{17}$ for large-scale convection,
${\cal R}_{\rm m}\sim 10^{10}$ for local turbulence)
of both PSSs and PNSs.
Unfortunately, the question of whether or not fast dynamo exist has
not been answered theoretically although many numerical and analytical
calculations strongly support the existence of kinematic fast dynamos
for given sufficiently complicated flows (Soward 1994, Childress \&
Gilbert 1995).
It is worth studying fluids without magnetic diffusion since the
diffusion timescale is much longer than the advection timescale,
$\tau_{\rm ohm}\gg \tau_{\rm v}$, for fast dynamos. The complex
flows, such as stretch-twist-fold, may effectively amplify the
field in this case.
The amplified strong magnetic fields are concentrated in filaments
with radii $l_{\rm f}$, which can be estimated to be
$l_{\rm f} \sim {\cal R}_{\rm m}^{-1/2}{\check L}
\sim 0.01/\sqrt{{\cal R}_{\rm m16}}$
mm (${\cal R}_{\rm m}={\cal R}_{\rm m16}\times 10^{16}$) by equating
the diffusion timescale of filament field, $l_{\rm f}^2/\kappa_{\rm m}$,
to the advection timescale $\tau_{\rm v}$.
Fast dynamos in PNSs have been considered by Thompson \& Duncan
(TD93) who suggested PNS dynamos as the origins of pulsar magnetism.
As the fluid parameters (e.g., ${\cal R}_{\rm m}$, $\nu$, $\kappa$) of
both PSSs and PNSs are similar, fast dynamos may also work for newborn
strange stars.

There are three timescales in the fast dynamo of PSSs: the diffuse
timescale $\tau_{\rm ohm}={\check L}^2/\kappa_{\rm m}$, the advection
timescale $\tau_{\rm v}={\check L}/{\check U}$, and the buoyancy
timescale $\tau_{\rm b}$,
\begin{equation}
\tau_{\rm b}\approx 4c\sqrt{\pi\over 3GM}RL^{1/2}\rho^{1/2}B^{-1}
\sim 10{\rm ms}~R_6L_5^{1/2}\rho_{15}^{1/2}B_{16}^{-1},
\end{equation}
($R=R_6\times 10^6$ cm, $L=L_5\times 10^5$ cm, $B=B_{16}\times 10^{16}$ G).
$B_{16}\sim 1.12~\rho_{15}^{1/2}v_8$ ($v=v_8\times 10^8$ cm
s$^{-1}$) if we assume that kinetic and magnetic energy densities
are in equilibrium. Filaments can be generated inside the
convective layer since turbulent convection is violent on small
scales ($\sim 100$ cm). $\tau_{\rm v}\la \tau_{\rm b}\ll \tau_{\rm
ohm}$, which means that filaments rise to the stellar surface by
magnetic buoyancy as soon as strong fields are created by fast
dynamos.
Magnetic buoyancy flow or convection could create and amplify the
poloidal (transverse) field from the azimuthal (aligned) field,
at the same time, the aligned field could also be created and
amplified from the transverse field by differential rotation as
discussed in section 2.1. Thus these processes may complete a
``dynamo cycle'' in newborn pulsars.

Let's give some estimates for fast dynamos. There are mainly two
types of flows in PSSs: eddy (convection) and shear (differential
rotation). Actually, these flows are coupled. However, we may
deal with them separately in order to have an overview of the
field generation processes.
For pure straining motion with velocity field ${\cal V}$,
\begin{equation}
{\cal V} = \tau_{\rm v}^{-1} (-x,y,0),
\end{equation}
which represents eddy flow to some extent, the magnetic field can
be amplified considerably to $B_{\rm max}^{\rm e}=B_0{\cal
R}_{\rm m}^{1/2}$ at a timescale $t_{\rm max}^{\rm e}={1\over
2}\tau_{\rm v}{\rm ln}{\cal R}_{\rm m}^{1/2}$ (Soward 1994). For
an initial field strength $B_0\sim 10^{10}$ G, $B_{\rm max}^{\rm
e}\sim 10^{18}{\cal R}_{\rm m16}^{1/2}$ G and $t_{\rm max}^{\rm
e}\sim 18 \tau_{\rm v3}{\rm ln}{\cal R}_{\rm m16}^{1/2}$ ms
($\tau_{\rm v}=\tau_{\rm v3}\times 10^{-3}$ s).
On another hand, the linear shear flow (Soward 1994),
\begin{equation}
{\cal V} = T_{\rm v}^{-1} (y,0,0),
\end{equation}
can amplify the field to
$B_{\rm max}^{\rm s}\sim B_0{\cal R}_{\rm m}^{1/3}
\sim 2\times 10^{15}{\cal R}_{\rm m16}^{1/3}$ G
at a growth time
$t_{\rm max}^{\rm s}\sim T_{\rm v}{\cal R}_{\rm m}^{1/3}
\sim 21T_{\rm v4}{\cal R}_{\rm m16}^{1/3}$ s
for $B_0=10^{10}$ G
($T_{\rm v}=10^{-4}T_{\rm v4}\sim {L/(V(R)-V(R-L))}$, see Fig.\ref{Fig.Dv}).
The timescale for field generation is thus from
$10^{-5}$ s (for ${\cal R}_{\rm m}\sim 10^{10}$)
to 10 s (for ${\cal R}_{\rm m}\sim 10^{16}$),
while the amplified fields could be
$10^{12}$ (for ${\cal R}_{\rm m}\sim 10^{10}$) to $10^{18}$
(for ${\cal R}_{\rm m}\sim 10^{16}$) G in filaments.
The magnetic fields emerging from under the stellar surface are likely to be
much smaller than $B_{\rm max}$ in filaments because the filaments
will expand to an approximately homogeneous field near and above
the surface owing to magnetic pressure.
Incoherent or coherent superposition of this buoyant field flux
might result in the observed dipole moment of pulsars (TD93).

Differential rotation may play an essential role for the
generation of large-scale magnetic fields through $\alpha-\Omega$
dynamo process, but even in the absence of differential rotation,
large-scale magnetic fields may be created. Owing to the alignment
of small-scale convection rolls parallel to the axis of rotation,
global magnetic fields can be generated as has been shown in
various dynamo models (see, for instance, Busse 1975).


\section{Any differences of dynamo-originated fields?}

As discussed in the previous section, magnetic dipole fields
stronger than $10^{11}$ G could be generated by dynamos in PSSs.
However, if pulsars are born as strange stars, an acute question
is how to explain the fields of millisecond pulsars with $\sim
10^8$ G since fields don't decay in strange stars with CSC.
Millisecond pulsars are supposed to be spun-up (``recycled'') as
the result of accretion from companion stars in their histories.
They could be strange stars with $\sim 10^{-5}M_\odot$ crusts,
although a lot of ordinary pulsars might be strange stars with
bare polar caps and much thinner crusts (Xu et al. 2001). We
suggest that accretion may reduce the dipole field strength from
$\sim 10^{12}$ G to $\sim 10^8$ G of strange stars in four
possible ways: (1) the accreted matter may screen or bury the
original fields (Bisnovati-Kogan \& Komberg 1974, Zhang et al.
1994, Zhang 2000); (2) field decays in the accretion-heated crust
(Konar \& Bhattacharya 1997); (3) the large scale fields (e.g.,
dipole structure) could be changed significantly because the
plasma accreted onto the polar caps would squeeze the frozen
magnetic fields toward the equator (Cheng \& Zhang 1998), and the
dipole field thus appears to decay; (4) magnetic fields frozen in
the bottom crust may dissipate and be annihilated during the
combustion process of the bottom matter into SQM when the crust
becomes too heavy to be supported by the Coulomb barrier.

A notable difference between strange stars and neutron stars is
that the magnetic field of a strange core is stable and does not
decay due to CSC, while the field of a neutron star should be
expelled from the interior to the crust where the Ohmic
dissipation occurs.
In present models for neutron star magnetism, the fields permeate
either the whole star or only the crust (see, e.g., a short review
by Mitra et al. 1999).
It is found by observational and statistical analyses that a
pulsar's field can only decrease substantially in the
accretion-phase but does not decay significantly during the
pulsar lifetime (Hartman et al. 1997).
New calculations of the timescale for Ohmic dissipation in the
crust have shown that the field can persist for more than
$10^{10}$ years (Sang \& Chanmugam 1987). It thus may be difficult
to distinguish the field evolutions of strange stars and neutron
stars in the radio pulsar-phase.
Nevertheless, the field decay modes in the accretion-phase could
be quite different for neutron stars and strange stars. In fact
the items (1)-(3) in the above proposals for field decay are
relevant to both neutron stars and strange stars, but item (4)
can only apply to strange stars. Furthermore, items (1)-(3) could
result in different physical processes for strange star and
neutron star because they have very different structures. For
example, as matter accretes, the radius of a strange star
increases, while the radius of a neutron star decreases. Also,
the material in neutron star crust moves continuously, and the
movement just pushes and squeezes the original field. However for
strange stars, a SQM phase-transition occurs when the accreted
material moves across the Coulomb barrier with thickness of $\sim
200$ fm. Actually, this phase-transition process has been
included in the explanation of some burst phenomena, such as
bursting X-ray pulsar GRO J1744-28 (Cheng et al. 1998), but the
field evolution in accreting strange star has not yet been
discussed extensively. We propose here that strange stars and
neutron stars may be distinguished by their field decays during
the accretion-phases. Future investigations of this issue will be
of interest.

As addressed in the previous sections, many fluid parameters for
the large-scale convection scenario in PSSs and PNSs are similar.
Both kind of stars have convective layers with thickness $\sim
10^5$ cm and flow velocities $\sim 10^8$ cm/s. The scale length of
turbulent eddies of both PSS and PNS is $\sim 2$ m since neutrino
viscosity damps flow only on scale greater than $l\sim l^{\rm
N}\sim 2$ m. The neutrino viscosities and the magnetic Reynolds
numbers in PSSs and PNSs are also not quite different: $\nu \sim
10^{10}$ cm$^2$/s, $\nu^{\rm N} \sim 10^8$ cm$^2$/s; ${\cal
R}_{\rm m}\sim 10^{16}$, ${\cal R}_{\rm m}^{\rm N}\sim 10^{17}$.
Therefore, the general configurations and strength of
dynamo-generated magnetic fields in PSSs and PNSs may be similar.
However, the fluid properties calculated in this paper for SQM are
rather uncertain since our knowledge of SQM fluid is less than
that of nuclear matter. Detailed studies of the fluid properties
in PSSs, such as the neutrino fraction and viscosity, are
necessary and may help us to see
 differences in the dynamo-originated fields.
In addition, the timescale available for dynamo action in PSSs may be
significantly smaller than that in PNSs since the energy gap for
proton superconductivity is order of 1 MeV (rather than 10-100 MeV
in SQM). This property may also affect the structure of the fields.


\section{Conclusion and discussion }

Pulsars could be neutron stars and/or strange stars. The origin
of neutron star magnetic fields has been discussed in literature,
while few papers are concerned with the origin of fields in strange
stars. In this paper, we have investigated the dynamo action in PSSs,
and have suggested that strange stars can have
magnetic fields of dynamo origin
similar to those in neutron stars. Our main conclusions
are as follows.

1. A significant fraction of gravitational energy has to be
converted to differential rotation energy if the angular momentum
of each mass element is conserved in the core collapse. Assuming
an initial {\em rotating} core to be approximated by the model of
Bruenn (1985), we have calculated the rotation velocity, the
velocity derivative, and the differential rotation energy of a
nascent strange star.

2. It is found that PSSs may have a convective layer of
thickness $\sim 2$ km in the outer part. The large-scale
convection has a velocity $\sim 10^8$ cm/s, while  local
turbulent eddies have a scale of $\sim 1$ m and a velocity of $\sim
10^5$ cm/s.

3. The energy density of differential rotation could be larger than
the turbulent energy density if the pulsar initial period $P \la
10$ ms. Assuming that most of the differential rotation energy is
converted to magnetic energy by dynamo action, we obtain the
dipole field strength as a function of $P$ and the pulsar period
evolution due to magnetic dipole radiation.

4. The fields ($10^{12}-10^{18}$ G) amplified by fast dynamo
action are concentrated in filaments with initial radii $\sim
0.01$ mm - $\sim 1$ cm and with growth times $10^{-5}$ s to 10 s.

5. Strange stars and neutron stars are expected to have different
accretion-induced field decay processes, which could be used to
distinguish them in the future.

6. Convection with the large scale $L$ is less likely to exist in
PSSs than in PNSs.

It is currently believed that anomalous X-ray pulsars (AXPs) and
soft gamma repeaters (SGRs) are magnetism-powered in the
``magnetar'' model (Duncan \& Thompson 1992, Thompson \& Duncan
1995). The magnetic reconnection near the surface causes the
emission in AXP and SGR.
Duncan \& Thompson (1992, TD93) suggested that the key parameter
that determines whether a PNS becomes an ordinary pulsar or a
magnetar is the Rossby number. {\em If} magnetars are strange
stars, we propose that, besides this key parameter, the initial
temperature and the density of trapped neutrinos can also affect
the formation of a magnetar since dynamo action could not be very
effective when CSC appears. Strange stars with very strong
magnetic field may act as ``magnetars'' since they could have
similar differentiated structure as neutron stars (Benvenuto et
al., 1990). However the neutron magnetar model faces a crisis
indicated by P\'erez Martinez et al.(2000) in that neutron stars
may undergo a transverse collapse if their fields exceed a
critical value. But for strange magnetars, a relatively small
magnetic momentum and large chemical potential of free quarks may
favor the formation of very high fields, although a further
investigation is
 needed to find the critical field strength beyond which a
strange star cannot be sustained against transverse collapse.

This paper may have two implications for the studies of the
$r-$mode instability (Anderson 1998; Madsen 1998, 2000; Lindblom
et al. 1998).
(1) As discussed in section 2.1, PSSs and PNSs should rotate
strongly differentially (particularly for $P\la 10$ ms). But the
numerical studies that have appeared in the literature are for
stars with {\em uniform} rotation. Certainly, it is reasonable to
assume that the result for uniform rotation is qualitatively
representative also for the case of differential rotation.
(2) The dynamo-originated strong magnetic fields inside strange
stars or neutron stars should be included in investigations of
the $r-$mode instabilities in pulsars. Although a very short
period $P$ will be favourable for the occurrence of $r-$mode
instability, a smaller $P$ would also be favourable for the
generation of a stronger magnetic field by dynamo action, thus
exerting a stabilizing influence. As the temperature decreases,
the bulk viscosity also decreases and the $r-$mode instability
may occur if there are no other strong dissipative effects.
Previously, Rezzolla et al. (2000) have shown that the
interaction of $r-$mode oscillations with the magnetic field
could be important and that the oscillations will be inhibited
 {\em if} the field is of the order $\sim 10^{16}
(\Omega/\Omega_{\rm B})$ G, where $\Omega$ is
 the angular velocity of the star and $\Omega_{\rm B}$ is the
value at which mass shedding
occurs.
Since the  fields of PSSs (or PNSs) depend on the
 nature of turbulence and on the rotation frequency,
 further study is needed to see whether
dynamo-generated fields can affect significantly the $r-$mode
oscillation.


\vspace{2mm}

{\em Acknowledgments.}
This research was supported by National Nature Sciences Foundation
of China (19803001), and by the Special Funds for Major State
Basic Research Projects of China.

\end{document}